\documentclass[conference]{IEEEtran}

\usepackage[T1]{fontenc}
\usepackage[utf8]{inputenc}
\usepackage{graphicx}
\usepackage{xcolor}
\usepackage{amssymb, amsmath}
\usepackage{tgtermes}
\usepackage{epstopdf}

\begin{document}

\title{Hybrid Beamforming for Massive MIMO Backhaul}

\author{ \IEEEauthorblockN{Namal Rajatheva and Elvino S. Sousa}
\IEEEauthorblockA{\\Department of Electrical and Computer Engineering \\
University of Toronto, Toronto, Ontario, Canada\\
Email: \texttt{\{namal.rajatheva,es.sousa\}@utoronto.ca}}
}

\maketitle

\begin{abstract}
The uplink where both the transmitter and receiver can use a large antenna array is considered. This is proposed as a method of antenna offloading and connecting small cell access points (SCAP) in a Two-Tier cellular network. Due to having a limited number of RF-chains, hybrid beamformers are designed where phase-only processing is done at the RF-band, followed by digital processing at the baseband. The proposed receiver is a row combiner that clusters sufficiently correlated antenna elements, and its' performance is compared against random projection via a Discrete Fourier Transform (DFT) matrix. The analogue to the row combiner is a column spreader, which is dependent on the transmit correlation, and repeats the transmitted signal over antenna elements that are correlated. A key benefit of this approach is to reduce the number of phase shifters used, while outperforming the DFT scheme. When only the transmitter has correlation and is RF-chain limited, the baseband precoding vectors are shown to be the eigenvectors of the effective transmit correlation matrix. Depending on the channel correlation, this matrix can be approximated to have a tridiagonal Toeplitz structure with the proposed column spreader (CS). The resulting eigenvalues have a closed form solution which allows us to characterize the sum rate of the system. Most interestingly, the associated eigenvectors do not require knowledge of the effective transmit correlation matrix to be calculated using an Eigenvalue Decomposition (EVD) method.  
\end{abstract}


\section{Introduction}
 \par In a Massive MIMO system we have links with a large number of antenna elements at the base station (BS), and a relatively small number at the user equipment (UE) due to size and power constraints. Recently, LTE Rel 13 proposed employing up to 64 antennas at the BS \cite{rel13} to increase network performance and meet performance requirements expected of future 5G networks. However, we can significantly improve the link capacity by including a large number of antenna elements close to the terminal,  and then focusing on the first link between the BS and the active relay (i.e. advanced UE). This is a MIMO (multiple input multiple output) channel where both $N_{r}$ and $N_{t}$ are large and different scenarios for the deployment of this advanced UE are possible. Within small cell networks, one application is to form a wireless backhaul \cite{wirelessbackhaul, wirelessbackhaul5G} that ensures connectivity for all small cell access points (SCAP). This removes the need of installing dedicated optical backhaul connections that are costly to network operators. In order to realize this, we characterize the performance of a wireless system where both the transmitter and receiver can employ Massive MIMO and the impact channel correlation can have. 
\par One of the bottleneck of Massive MIMO systems is the RF (radio frequency) circuitry required for hardware implementation. Each antenna element typically has an associated RF-chain, which is responsible for converting the received modulated signal to the baseband (BB) and sampling it appropriately so that the received analog signal is converted to a digital signal and traditional digital signal processing (DSP) techniques can be applied for signal estimation and symbol detection. There needs to changes made to the hardware architecture if we are to implement transceivers with $\sim$100 antenna elements as is being considered in ongoing standardization work. Hence it has been proposed to use a limited RF-chain system to overcome this practical constraint. 
\begin{figure}[!h]
 \begin{center}
   \includegraphics[width=0.5\textwidth]{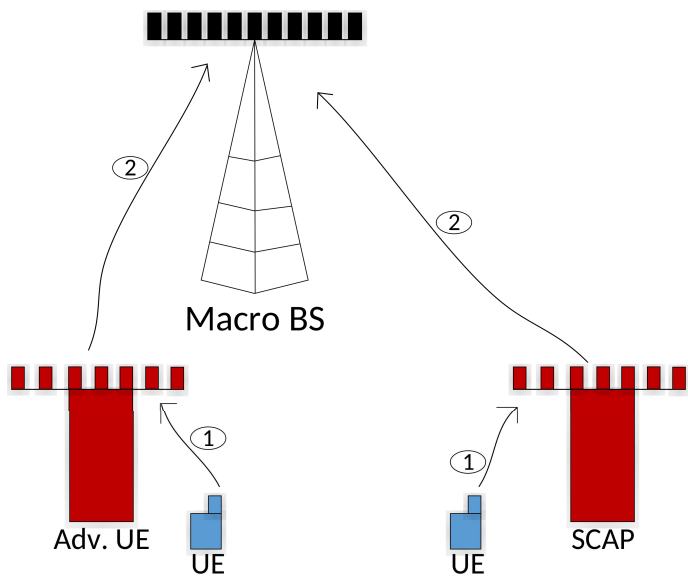}
\end{center}
 \caption{Possible scenarios for Massive MIMO deployment in Two-Tier Cellular Networks}
 \label{fig:envi}
\end{figure}
The limited RF-chain problem can be formulated as the design of a RF-band processing matrix that is in cascade with the baseband matrix, also called hybrid beamforming. The RF-band matrix consists of phase only elements with fixed amplitude that can be implemented by a network of variable phase shifters. Equal gain combining (EGC) is a form of phase only processing without the need of amplitude control, and has been shown to approach the performance of the matched filter (MF) for the single user case. Another extensively studied approach is to select a subset of antenna elements which equals the number of RF-chains available. The antenna selection problem results in searching over all possible combinations that optimizes some utility function, such as maximizing the sum rate. When there are a large number of antenna elements, an exhaustive search is not possible and heuristic methods can be applied. One such method is to look at the effect of adding an antenna to the channel capacity formula \cite{capacityselection} and iteratively choosing antenna elements which give the highest increase. Within the context of spatial beamforming, adaptive nulling with no amplitude control has been considered in \cite{optimumphase} and related works, where the beamforming weights are found by solving an optimization problem through an iterative algorithm. Several recent papers have considered minimizing the Frobenius norm of the error matrix with respect to the fully digital solution\cite{hybridnorm}. In \cite{rfchain} and related works, a Discrete Fourier Transform (DFT) matrix is considered for the RF-processing stage due to its relative ease of implementation and simplifying the precoder/postcoder design problem. Similarly in \cite{egchybrid}, rather than jointly designing the RF and baseband filters, the baseband filter is designed first and the RF filter is calculated using an iterative algorithm. Other works have interpreted the limited RF-chain system in relation to an antenna switch architecture\cite{selectionmassivemimo}, which gives us more flexibility in designing the RF-band phase only matrices. Going along this line of reasoning, a simple row/column combiner is proposed for the RF-processing stage to take advantage of the channel correlation. This arises in a wireless system due to antenna geometry, operating frequency and the environments propagation characteristics. The performance of this scheme is compared against employing a random subspace projection through a DFT matrix, and designing digital baseband precoding matrices with the effective channel. Hadamard matrices were another random projection that was considered, but consistently underperformed compared to employing a DFT projection. 
\par The contributions of this paper are as follows. For the limited RF-chain system, a row combiner is proposed as the RF filter. This relaxes the constraint of phase-only elements  at the RF band to include zero gain elements, and reduces the number of phase shifters required to be proportional to the number of independent streams. In fact, the RF filter is dependent on channel correlation, and does not require channel state information (CSI). When the transmitter is RF-chain limited, the same signal is repeated over transmit antenna elements which are sufficiently correlated. When only the transmitter has correlation and the receiver has Massive MIMO (i.e. base station in uplink), choosing an appropriate cluster size for the transmit RF filter results in closed form evaluation of the sum rate and precoding vectors. This requires channel correlation information at the transmitter (CCIT), which incurs less overhead than feeding back CSI. 
\subsection{Background and Notation}
Let $N_{t}$ be the number of antenna elements and $L_{t}$ the number of RF-chains at the transmitter. The precoding matrix is the cascade of the RF and baseband (BB) matrix
\begin{equation}
\bold{V} = \bold{V}_{RF}\bold{V}_{BB}\label{eq:hybrid}
\end{equation}
where $\bold{V}_{RF}$ is an $N_{t}\times L_{t}$ matrix and $\bold{V}_{BB}$ is a $L_{t}\times d$ matrix, where $d$ is the number of independent streams to be transmitted. The case where $L_{t}=N_{t}$ corresponds to the traditional case where precoding techniques for the multiple transmit antenna case can be applied. The elements of $\bold{V_{BB}}$ have no constraints, while the elements of $\bold{V_{RF}}$ is given by $a_{ij}e^{\phi_{ij}}$, where $a_{ij}$ can be either 0 or some fixed constant. This relaxes the constraints that is typically considered when designing hybrid beamforming vectors. Similarly at the receiver we can have $N_{r}$ and $L_{r}$, with the postcoder given by $\bold{W}^{H}=\bold{W}_{BB}^{H}\bold{W}_{RF}^{H}$, where $(\cdot)^{H}$ is used to denote the Hermitian of a matrix. The traditional approach to limited RF-chain systems is to select a subset of antenna elements and results in the RF filter being all zeros except at the index of the antenna elements selected. An example of the RF filter is given in Eq. \ref{eq:antsel} for which signal from antenna elements with index 1,5,7 are chosen when the number of RF-chains is 3
\begin{equation}
\bold{W}_{RF} = \begin{pmatrix} \bold{e}_{1} & \bold{e}_{5}  & \bold{e}_{7} \end{pmatrix} \label{eq:antsel}
\end{equation}
 where $\bold{e}_{i}$ is a column vector which is zero everywhere except at the $i$-th index, where it contains a 1. For the case where $\bold{W}_{RF}$ allows phase only elements of the form $e^{j\phi_{m,n}}$, each RF-chain has associated with it $N_{r}$ variable phase shifters with which to weight the received signal.
\begin{figure}[!h]
 \begin{center}
   \includegraphics[width=0.5\textwidth]{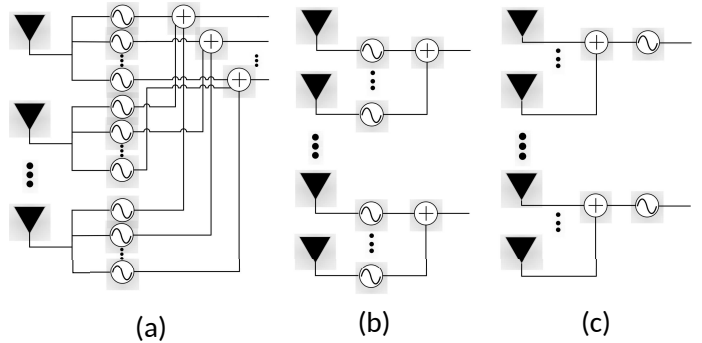}
\end{center}
 \caption{Different RF architecture with limited number of RF-chains: (a) Each antenna element feeds different phase shifted version of its signal to each RF-chain (b) Cluster of antenna elements are associated with a RF-chain, and the phase for each can be controlled (c) Signal from cluster of antenna elements are added and fed to RF-chain with no phase processing, similar to network switch architecture}
 \label{fig:rf_chain}
\end{figure}
While the DFT requires no knowledge of the transmit correlation at the RF stage, it requires having $N_{t}L_{t}$ phase shifters, while the former only requires at most $N_{t}$. If the relative phase between clusters is controlled, then this number can be reduced to $L_{t}$, while if no control is required only 1 phase shifter is used at the RF stage to act as a reference.

\section{System Model}
Consider a single cell where a user is transmitting with $N_{t}<N_{r}$ antenna elements under flat fading conditions. The received signal can be written as
\begin{equation}
\bold{y} = \bold{H}\bold{x} + \bold{n}
\end{equation}
where $\bold{H}$ is the channel gain matrix. Under the Kronecker model for correlation, the channel gain matrix can be modelled as $\bold{R}_{r}^{\frac{1}{2}}\bold{G}\bold{R}_{t}^{\frac{1}{2}H}$, while $\bold{n}$ is additive white Gaussian noise. $\bold{G}$ is a $N_{r}\times N_{t}$ matrix whose elements are i.i.d. and follow distribution $\mathcal{CN}(0,1/N_{r})$ and $\bold{R}_{t}^{\frac{1}{2}}\bold{R}_{t}^{\frac{1}{2}H}=\bold{R}_{t}$,  $\bold{R}_{r}^{\frac{1}{2}}\bold{R}_{r}^{\frac{1}{2}H}=\bold{R}_{r}$. $\bold{R}_{r}$ and $\bold{R}_{t}$ are the correlation matrices at the receiver and transmitter with size $N_{r}\times N_{r}$ and $N_{t}\times N_{t}$ respectively and correlation between the $i,j$-th antenna elements is $\alpha^{\left(i-j\right)^{2}}$ for a uniform linear array antenna \cite{correl}. The transmitted vector symbol $\bold{x}$ is given by
\begin{equation}
\bold{x} = \bold{V_{RF}}\bold{V_{BB}}\bold{s}
\end{equation}
where $\bold{s}$ is the $L_{t}\times1$ information symbol vector to be transmitted. Normalized precoding is employed to ensure  the average transmit power constraint $\mathbb{E}[||\bold{x}||^2]\leq P$ is satisfied. 
\section{Designing Hybrid Beamformers}
\subsection{Limited RF-chain at Receiver only, $L_{r} \leq N_{r}$}
For an asymptotically large number of antenna elements at the receiver, the MF would maximize the signal to noise ratio (SNR) while suppressing the inter-stream interference, assuming no correlation. When there is RF-processing involved, care must be taken with respect the impact of noise colouring as it may further complicate the processing required at the baseband. One natural choice for the RF filter is to take the phase of each element in $\bold{H}$ and form the phase gain matrix $\bold{H}_{\theta}$ whose $\left(m,n \right)$-th element is given by $e^{j\theta_{n}(m)}$. For the $i$-th transmitted stream, the SNR can be found by writing out 
\begin{align}
\frac{1}{N_{r}}\bold{w}_{i}^{H}\bold{y} = \frac{1}{N_{r}}\sum_{l=1}^{N_{r}}|\bold{h}_{i}(l)|s_{i} + \frac{1}{N_{r}}\sum_{j\neq i}\bold{w}_{i}^{H}\bold{h}_{j} +\tilde{n}
\end{align}
$\tilde{n}$ being the effective noise with variance $N_{r}^{-1}$. Using the weak law of large numbers, we can approximate each term as follows
\begin{align}
\frac{1}{N_{r}}\sum_{l=1}^{N_{r}}|\bold{h}_{i}(l)| &\rightarrow \mathbb{E}[|\bold{h}_{i}(l)|]\\
\frac{1}{N_{r}}\bold{w}_{i}^{H}\bold{h}_{k} &= \frac{1}{N_{r}}\sum_{l=1}^{N_{r}} \bold{h}_{k}(l)e^{-j\theta_{i}(l)} \nonumber\\
              &\rightarrow \mathbb{E}[\bold{h}_{k}(l)e^{-j\theta_{i}(l)}]=0\label{eq:indep}			 
\end{align}
assuming $N_{r}$ grows asymptotically large. Eq. \ref{eq:indep} follows from the independence of variables indexed by $i,k$ and under the assumption $\bold{h}_{k}(l)$ is complex Gaussian. Therefore applying the EGC at the receiver with an asymptotically large number of receive antenna elements suppresses the inter-stream interference. The resulting SNR is given by
\begin{equation}
\gamma_{i} =  N_{r}\mathbb{E}[|\bold{h}_{i}(l)|]^{2} 
\end{equation}
Additionally, it follows that the noise correlation matrix can be approximated by $\bold{W}_{RF}^{H}\bold{W}_{RF}\rightarrow N_{r}\bold{I}_{L_{r}}$. In practice $N_{r}$ cannot be made asymptotically harge, hence further processing is required at the baseband. One straightforward digital baseband filter is to simply invert the effective channel \cite{lowcomplhybrid}, hence the received filter can be written as follows
\begin{equation}
\bold{W}^{H} = \left( \bold{H}^{H}_{\theta}\bold{H} \right)^{-1}\bold{H}^{H}_{\theta}
\end{equation}
Now consider the case where there is correlation at the receiver side. The correlation is induced amongst the elements of the columns of $\bold{H}$, hence the RF filter at the receiver side must appropriately combine the rows. In the extreme case when there are disjoint groups of highly correlated rows, each group would be added together constructively while ignoring the other rows. For moderate values of $\alpha_{r}$ corresponding to receiver correlation of neighbouring antenna elements, $\bold{R}_{r}$ would have a sparse structure for large $N_{r}$. This leads to the idea of using a simple row combiner at RF filter, given by 
\begin{equation}
\bold{W}_{RF}^{H} =  \frac{1}{\sqrt{K}}\begin{pmatrix} \bold{1}_{1\times K} & \bold{0}_{1\times K} & \cdots & \bold{0}_{1\times K}\\ \bold{0}_{1\times K} & \bold{1}_{1\times K} & \cdots & \bold{0}_{1\times K}\\  \vdots   & \vdots & \ddots & \vdots \\ \bold{0}_{1\times K}  & \bold{0}_{1\times K} & \cdots & \bold{1}_{1\times K}\end{pmatrix}\label{eq:rf_receive}
\end{equation}
and assumes $N_{r}=KL_{r}$. The parameter $K$ can be interpreted as a correlation interval of the antenna array, where for a fixed $N_{r}$ a higher $K$ would lead to supporting fewer independent streams that will be decoded at the baseband. Assuming the number of transmitted streams is equal to the number of RF-chains, this leads to an architecture where all the diversity gain is achieved at the RF band, and the baseband is responsible for the multiplexing gain. Typically correlation arises due to antenna spacing constraints, and as a general rule of thumb half wavelength spacing is used. It has been shown that there is some merit in using additional antenna elements, at the cost of increasing the correlation \cite{transmitcorrelation, corrmassive}. Ideally, there would $L_{r}$ disjoint sets of antenna elements which undergoes RF processing. One specific case of this is antenna selection, where each set only contains one antenna element and the baseband would perform the traditional MIMO techniques. This leads to limited performance gains compared to employing phase only processing, as will be seen in the performance analysis section. For simplicity, taking $N_{r}$ to be fixed and $\alpha_{r}$ to be a parameter of the environment, the strategy would be use an appropriate value for $K$, which determines the number of independent streams. Some papers have considered the approach of dynamically allocating the number of RF-chains for limited RF-chain system \cite{rfchain}, and presents another level of optimization that be can carried out by the base station scheduler. Since $\alpha_{r}$ is limited to take on values between 0 and 1, one choice for $K$ can be obtained by ensuring the condition
\begin{equation}
\alpha_{r}^{\left(\frac{K}{2}+1\right)^{2}} \leq \epsilon
\end{equation}  
holds for some threshold $\epsilon$ that is close to zero. For one cluster of antenna elements, this is the correlation between the center element and edge element of the adjacent cluster. The smallest $K$ which satisfies this constraint is denoted by $K_{c}$ and can be calculated by
\begin{equation}
K_{c} = 2\left \lfloor          \sqrt{-\frac{\log\epsilon}{\log\alpha_{r}}}       \right \rfloor \label{eq:cons}
\end{equation}
A conservative value of $\epsilon$ can be 0.1, which for practical purposes can correspond to being uncorrelated. For a system which has parameter values of $N_{r}=256$ and $\alpha_{r}=0.8$ yields a value of $K=8$, which results in $L_{r}=32$ and the resolvable number of transmit streams is $\min \left(L_{r}, N_{t} \right)$. Fixing the number of receive antenna elements to be $N_{r}$, the capacity of the system can firstly be improved by $N_{t}$, but it is known that there is diminishing returns as $N_{t}$ approaches $N_{r}$. The base case is taken to be when $N_{t}=N_{r}$, which corresponds to both the transmitter and receiver having full flexibility and not limited by the number of RF-chains. Using the relation in Eq. \ref{eq:cons} results in reducing the number of RF-chains at the receiver, at the cost of reducing the sum rate. However, at high SNR the performance gap between the limited RF-chain system and the traditional system was observed to shrink.
\subsection{Limited RF-chains at Transmitter}
Now suppose there is only correlation at the transmitter side. The strategy is similar to what was done for the receiver in the prior section. The channel gain matrix can be written as
\begin{equation}
\bold{H} = \bold{G}\bold{R}_{t}^{\frac{1}{2}H}
\end{equation}
 which induces a correlation amongst the elements of each row, i.e. across the columns. Geometrically, the columns of $\bold{H}$ which are adjacent can be taken to be vectors pointing in the same direction, depending on the correlation parameter $\alpha_{t}$ being sufficiently large enough and vice versa. The extreme case is where clusters of the columns of $\bold{H}$ are perfectly correlated while inter-cluster elements have no correlation. Here the transmitter could simply repeat the signal over each element that is in the same cluster. Therefore, the transmit RF filter is given by
\begin{equation}   
\bold{V}_{RF} = \bold{W}_{RF} \label{eq:rf_spread}
\end{equation}
where $\bold{W}_{RF}$ is given by Eq. \ref{eq:rf_receive}. This filter performs the analogue task of spreading the signal over a sufficiently correlated portion of the channel, while the receiver was shown to combine signals that are sufficiently correlated. For this reason, we refer to RF filter given by Eq. \ref{eq:rf_spread} as a Column Spreader (CS). Assuming the Kronecker model where the correlation caused at the receiver is independent to the correlation at the transmitter, $\bold{W}_{RF}$ and $\bold{V}_{RF}$ can be used in tandem when both the transmitter and receiver are limited by the number of RF-chains. Ideally, depending on the channel gains the power of each repeated transmitted signal in one cluster can be controlled. However, this is not possible due to the constraint imposed by Eq. \ref{eq:hybrid}. \par A possible improvement is rather than simply adding correlated rows, to multiply each row by a phase such that the cluster of rows can be interpreted to be oriented to be in the same general direction. For two vectors $\bold{h}_{1}$ and $\bold{h}_{2}$, the problem can be stated as finding the phase $\phi$ such that the norm of $\bold{h}_{1}+ \bold{h}_{2}e^{j\phi}$ is maximized:
\begin{align}
\max_{\phi}||\bold{h}_{1}+ \bold{h}_{2}e^{j\phi}||^{2} &= \max_{\phi} \bold{h}_{1}^{H}\bold{h}_{2}e^{j\phi} + \bold{h}_{2}^{H}\bold{h}_{1}e^{-j\phi} \\
                                                                                              &= \max_{\phi} \Re{\left\{\bold{h}_{1}^{H}\bold{h}_{2}e^{j\phi} \right\}}
\end{align}
where $\Re$ denotes the real part and is achieved by letting $\phi$ be the phase of $\bold{h}_{2}^{H}\bold{h}_{1}$. For a general number of vectors there is no straightforward way to calculate the phases in one go. Instead we can calculate them iteratively by taking resulting vector, redo the calculation with it and $\bold{h}_{3}$ and so on. As with equal gain combining, combining with phase-only constraints can lead to performance degradation in instances where the norm of one vector is low due to the corresponding antenna element being in deep fade but has the same noise contribution. In \cite{Molisch}, a similar problem is considered where the phase-only combining vector is designed by first calculating the vector with elements of varying amplitude, taking the phase of each element and seeing which subset of phases maximizes the normalized inner product with the prior vector. This constraint arises in our case due to fixed transmit power, and affects the resulting SNR. \par In \cite{correlsnr} it was shown that to maximize the average SNR with transmit correlation, the precoding vector should be the eigenvectors of $\bold{R}_{t}$. When the receiver has a large number of receive antenna elements, this in fact leads to maximizing the instantaneous SNR. Suppose a single stream is to be transmitted, the received signal can be written as 
\begin{equation}
\bold{y} = \bold{G}\bold{R}_{t}^{\frac{1}{2}H}\bold{V}_{RF}\bold{v}_{BB}s_{1} +\bold{n} \label{eq:rec_rf}
\end{equation}
The matched filter maximizes the SNR, and is given by
\begin{align}
\gamma &= \bold{v}_{BB}^{H}\bold{V}_{RF}^{H}\bold{R}_{t}^{\frac{1}{2}}\bold{G}^{H}\bold{G}\bold{R}_{t}^{\frac{1}{2}H}\bold{V}_{RF}\bold{v}_{BB} \nonumber\\
           &\rightarrow  \bold{v}_{BB}^{H}\bold{V}_{RF}^{H}\bold{R}_{t}\bold{V}_{RF}\bold{v}_{BB} \label{eq:massive}\\
           &=  \bold{v}_{BB}^{H}\tilde{\bold{R}}_{t}\bold{v}_{BB} \label{eq:massive_corr}
\end{align}
where Eq. \ref{eq:massive} follows from applying the weak law of large numbers to approximate $\bold{G}^{H}\bold{G}$ as $\bold{I}_{N_{t}}$ and the receiver is taken to be correlation free. Hence it follows that $\bold{v}_{BB}$ is the eigenvector corresponding to the maximum eigenvalue of $\tilde{\bold{R}_{t}} =\bold{V}_{RF}^{H}\bold{R}_{t}\bold{V}_{RF} $. For multiple streams the eigenvectors of $\tilde{\bold{R}_{t}}$ can be used since they are orthogonal. The elements of $\bold{R}_{t}$ can be written as
\begin{equation}
\bold{R}_{t} = \begin{pmatrix}1 & \alpha _{t}& \alpha_{t}^{4} &\cdots& \alpha_{t}^{\left(N_{t}-1\right)^{2}} \\ \alpha_{t} & 1 & \alpha_{t} & \cdots & \alpha_{t}^{\left(N_{t}-2\right)^{2}} \\   \alpha_{t}^{4} & \alpha _{t}& 1 & \cdots & \alpha_{t}^{\left(N_{t}-3\right)^{2}} \\ \vdots & \vdots & \vdots & \ddots & \vdots  \end{pmatrix}
\end{equation}
Depending on the choice of $K_{c}$, $\tilde{\bold{R}}_{t}$ can be approximated as a symmetric tridiagonal Toeplitz matrix, for which the eigenvalues and eigenvectors are well known \cite{eigval}. Letting $a = \tilde{\bold{R}}_{t}\left(1,1\right)$ and $b = \tilde{\bold{R}}_{t}\left(1,2\right)$, the $i$-th eigenvalue is
\begin{align}
\lambda_{i} = a + 2b\cos\left(\frac{i\pi}{L_{t}+1}\right) \label{eq:eigenvalues}
\end{align}
 and associated eigenvector 
\begin{equation}
\bold{v}_{i} = \begin{bmatrix} \sin\left(\frac{i\pi}{L_{t}+1}\right) & \sin\left(\frac{2i\pi}{L_{t}+1}\right)& \cdots & \sin\left(\frac{L_{t}i\pi}{L_{t}+1}\right)  \end{bmatrix}^{T} \label{eq:eigenvectors}
\end{equation}
where $i=1,..,L_{t}$ and $\bold{v}_{i}$ is a column vector. Note that this vector does not depend on knowledge of $\bold{\tilde{R}}_{t}$, besides the fact that it is tridiagonal. The sum rate is then given by 
\begin{equation}
R_{sum} = \sum_{i=1}^{L_{t}}\log_{2}\left(1+\lambda_{i}P_{i}\right) \label{eq:sum_rate}
\end{equation}
which can be further optimized by employing the well known waterfilling algorithm for a fixed total power constraint. This results in selecting a specific subset of precoding vectors according to their corresponding eigenvalues. In the high SNR regime, equal power allocation suffices and the gain from employing waterfilling was observed to be minimal in simulation. The prior derivation assumed the receiver had full RF-chain capability and no correlation, which can correspond to uplink communication between a SCAP and BS.   
\section{Performance Analysis}
First only the transmit side is considered to have correlation and be RF-chain limited. The effective channel is denoted by $\tilde{\bold{H}} = \bold{H}\bold{V}_{RF}$ and its' capacity is calculated as
\begin{equation}
C = \log_{2}\left|\bold{I}_{N_{r}}+\tilde{\bold{H}}\bold{Q}\tilde{\bold{H}}^{H}\right|
\end{equation}
where $\left|\cdot\right|$ denotes the determinant and $\bold{Q}$ is the transmit source correlation matrix with trace constraint $\mathrm{Tr}\left( \bold{Q}\right) \leq P$ and waterfilling is employed for per-stream power allocation. The sum rate is calculated by forming a signal to interference plus noise ratio (SINR) after the normalized precoder and postcoder have been applied. The two RF filters are the column spreader (CS) defined in Eq. \ref{eq:rf_spread} and a random projection where the phase elements follow the structure of a DFT matrix. 
\begin{figure}[!h]
 \begin{center}
   \includegraphics[width=0.5\textwidth]{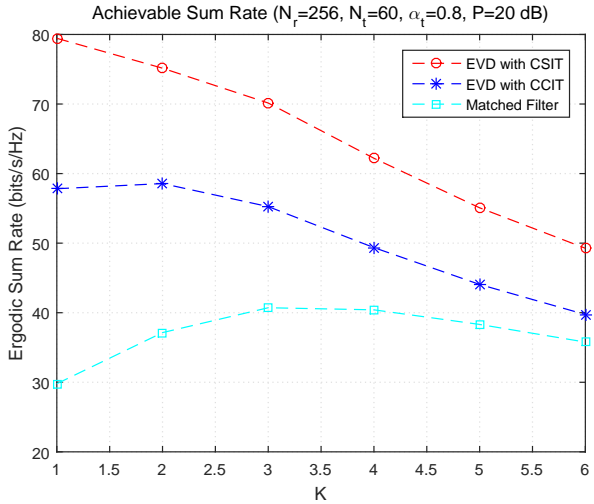}
\end{center}
 \caption{Achievable sum rate with CS as RF filter and corresponding baseband filter. CCIT assumes effective correlation matrix is tridiagonal. For massive MIMO effects to occur, we require $\sim$100's antenna elements at receiver with no correlation.}
 \label{fig:plot1}
\end{figure}
\par
\begin{figure}[!h]
 \begin{center}
   \includegraphics[width=0.5\textwidth]{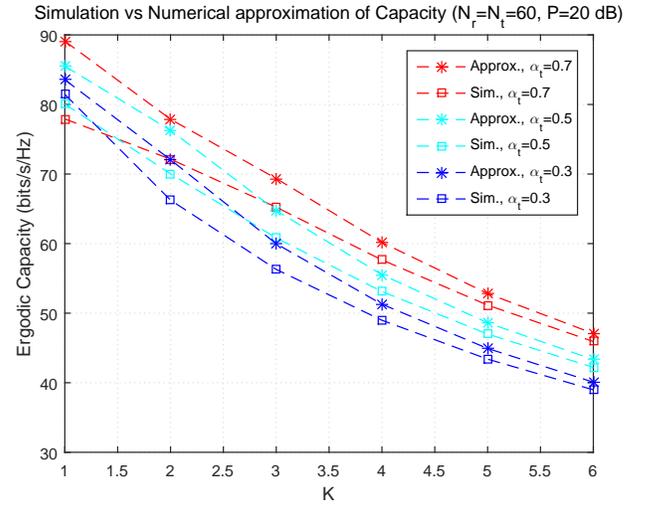}
\end{center}
 \caption{Effect of varying number of RF-chains when transmitter has correlation for fixed transmit power and using column spreader (CS) filter, $L_{t}=N_{t}/K$. Higher correlation requires higher values of $K$ for approximation to be valid, and results in better performance.}
 \label{fig:plot4}
\end{figure}

 In the $256\times 60$ MIMO system in Fig. \ref{fig:plot1}, using CCIT and taking the precoding vectors to be given by Eq. \ref{eq:eigenvectors} results in a constant gap performance in sum rate when compared to using full CSIT. For the matched filter, waterfilling was employed assuming there was no interference, effectively having the same overhead as EVD using CCIT. There are some diversity gains for the MF receiver when $K=3$, by taking advantage of the presence of transmit correlation which is not possible from an antenna selection scheme. Fig. \ref{fig:plot4} shows the capacity obtained with the column spreader (CS) filter using both numerical approximation given by Eq. \ref{eq:eigenvalues}, Eq. \ref{eq:sum_rate} and  carrying out Monte-Carlo simulations. As $K$ (defined in Eq. \ref{eq:cons}) increases for a fixed transmit correlation ($\alpha_{t}$), the approximation of $\tilde{\bold{R}}_{t}$ in Eq. \ref{eq:massive_corr} as a tridiagonal matrix is more applicable. At relatively high levels of SNR the approximation yields an upper bound on the capacity, and having a higher correlation leads to better performance with transmit power fixed at $20$ dB. It should be noted that less receiver antenna elements were required for characterization of the system through the eigenvalues of the effective channel matrix, as opposed to sum rate achieved by precoding using the eigenvectors in Eq. \ref{eq:eigenvectors}.
  \par
\begin{figure}[!h]
 \begin{center}
   \includegraphics[width=0.5\textwidth]{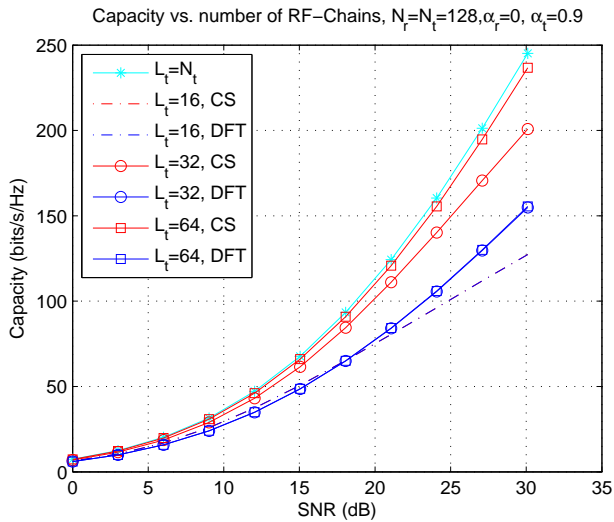}
\end{center}
 \caption{Transmitter has high channel correlation and is RF-chain limited. Using $25\%$ of RF chains with column spreader significantly outperforms random DFT scheme.}
 \label{fig:plot2}
\end{figure}

\begin{figure}[!h]
 \begin{center}
   \includegraphics[width=0.5\textwidth]{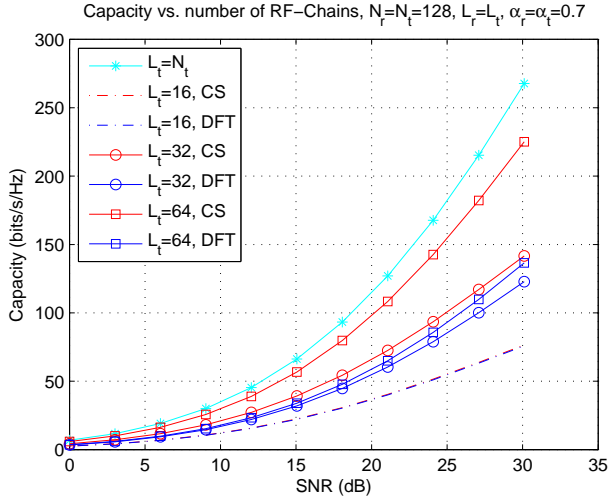}
\end{center}
 \caption{Both transmitter and receiver have channel correlation and are RF-chain limited. Adding correlation at the receiver results in a performance loss for both the DFT projection and column spreader, with the latter still edging out the former when using $L_{t}=32$.}
 \label{fig:plot3}
\end{figure}

 The column spreader performs better in the limited RF-chain regime when looking at the curves for $L_{t}=32$ in Fig. \ref{fig:plot2}. This comes despite the decrease in overall capacity ($L_{t}=N_{t}$). Using the DFT as a random projection performs comparably only when $L_{t}=16$, one-eighth of the number of antenna elements.
Rather than optimizing over the number of RF-chains $L_{t}$ to use, $K_{c}$ can be calculated using the heuristic rule according to Eq. \ref{eq:cons}. It was observed that just increasing the number of RF-chains leads to diminishing returns with respect to the sum rate in Fig. \ref{fig:plot4}. When the receiver is also RF-chain limited, the receive RF filter is $\bold{W}_{RF}^{H}=\bold{V}_{RF}^{H}$. The capacity is calculated by taking the effective channel to be $\tilde{\bold{H}}=\bold{V}_{RF}^{H}\bold{H}\bold{V}_{RF}^{H}$, and modifying the noise covariance matrix to be $\bold{I}_{L}$, where it has been assumed $L_{r}=L_{t}$. Reducing the correlation at the transmitter compared to that of Fig. \ref{fig:plot2}, and introducing correlation at receiver, it can be seen in Fig. \ref{fig:plot3} that for $L_{t}=32$ with the CS filter performs as good as the DFT filter that has $L_{t}=64$.  

\section{Conclusion}
We considered the scenario where both the transmitter and receiver can use a large number of antenna elements, to make full use of Massive MIMO gains. One such application is the connection of small cell access points through a wireless backhaul, rather than an optical backhaul. Due to hardware constraints it is necessary to consider Hybrid beamforming approaches which compromise between Massive MIMO gains and having a practically realizable system. When only the transmitter is RF-chain limited and has channel correlation, the receiver can use the matched filter to form an effective channel that is the transmit correlation matrix. The proposed RF-filter uses the correlation to form clusters of sufficiently correlated antenna elements. On the transmitter side, this results in spreading the signal while the receiver performs the opposite and combines the received signal. Using an appropriate cluster size, the effective transmit correlation matrix can be approximated as a tridiagonal Toeplitz matrix. This results in characterizing the system capacity by closed-form expression of the eigenvalues and corresponding eigenvectors as transmit precoding filters. The proposed column spreader (CS) and row combiner (RC) RF filters where shown to outperform a random phase only DFT projection, which requires significantly more phase shifters as outlined in Fig.  \ref{fig:rf_chain}. Future work should be done on characterizing the performance gains with different correlation models, and impact of mismatched CCI.
     
\bibliographystyle{ieeetr}

\bibliography{main}

\end{document}